\documentstyle{article}

\newcommand{\beq}{\begin{equation}}
\newcommand{\eeq}{\end{equation}}

\def\<{\langle}
\def\>{\rangle}
\newcommand{\complex}{{\kern .1em {\raise .47ex\hbox {$\scriptscriptstyle |$}}\kern -.4em {\rm C}}}
\newcommand{\real}{{{\rm I} \kern -.19em {\rm R}}}

\title{Quantum cryptography and long distance Bell experiments:
How to control decoherence
}

\author
{N. Gisin\\
J. Brendel, J-D. Gautier, B. Gisin, B. Huttner, \\ 
G. Ribordy, W. Tittel, H. Zbinden\\
{\protect\small\em Group of Applied Physics}\\
{\protect\small\em University of Geneva, 1211 Geneva 4, Switzerland}
}
\date{\today}

\begin{document}

\maketitle

\begin{abstract}
Several mechanisms that affect one and two photon coherence in optical
fibers and their remedies are discussed. The results are illustrated on
quantum cryptography experiments and on long distance Bell inequality tests. 
\end{abstract}


\section{Introduction}\label{int}
The implementation of 1- and 2-photon quantum communication protocols in
km long optical fibers suffers from several decoherence mechanisms. In this contribution
we review the main ones and illustrate how one can control them. 

Photons are characterized by three (non independent) parameters: 
their temporal coherence, their polarization and their frequency spectrum. In the next three
sections decoherence affecting each of these parameters are presented, together
with counter-measures. The first one, in the time domain,
leads to a useful measurement method of polarization mode dispersion. Mastering the second
one, depolarization, leads to a practical implementation of quantum cryptography. Finally,
the phenomenon of two-photon chromatic dispersion cancelling opens the route to
long distance Bell experiments.

\section{Polarization Mode Dispersion: Decoherence in the time domain}\label{PMD}
Real fibers are not perfectly circular. Consequently, the two polarization modes are
not degenerate and propagate at different phase and group velocities. The difference
in group velocities results in Polarization Mode Dispersion (PMD). The phenomenon
of PMD is presently a very severe limitation to high speed optical communication. In
addition to the presence of two group velocities, PMD is characterized by random polarization
mode coupling: some energy of the fast mode couples to the slow mode and vice-versa.
The locations where such couplings take place and their extend are very sensitive to thermal and mechanical
variations. Hence, in practice, the coupling is described as a random phenomenon
\cite{PMDJLWT,PMDCOST241}. The magnitude of the dispersion ranges from a few tenths of
a picosecond up to tens of picoseconds. Because of its stochastic nature, PMD is measured
in units of ps/$\sqrt{km}$. 

Direct measurement of PMD is a non trivial
task. When light with a short coherence time (typically light from a LED with
$\tau_c\approx 0.05$ ps) propagates down a fiber, the dispersion is larger than the
coherence, producing decoherence. However, coherence can be recovered by connecting
an interferometer at the end of the fiber, see figure 1. When the interferometer is
unbalanced, light that went out of coherence in the fiber by precisely the amount
of imbalance of the interferometer can be brought back into coherence. This leads
to interference fringes even when the interferometer's imbalance is larger than
the source coherence, see figure 2. This simple technique to recohere light is widely used by the 
telecom industry to measure PMD \cite{PMDBoulder}.

An interesting generalization using 2-photon interferometry was demonstrated by
A. Sergienko and A. Muller, see \cite{PMDSergienko,2photonBoulder98}.

\section{Depolarization: Decoherence in the polarization domain}\label{depol}
A single photon state $|\Psi_z\>$ at position $z$ along the fiber can be described as follows:
\beq
|\Psi_z\> = \int_0^\infty \psi_z(\omega) |1_\omega\> d\omega  
\eeq
where $|1_\omega\>$ denotes the 1-photon state at frequency $\omega$ and 
$\psi_z(\omega)\in \complex^2$ is a (non normalized) Jones vector describing the
polarization of the frequency component $\omega$ with the square norm $|\psi_z(\omega)|^2$   
the corresponding intensity. Let us introduce the Poincar\'e vectors:
\beq
\vec m_z(\omega)=\frac{\<\psi_z(\omega)|\vec\sigma|\psi_z(\omega)\>}{\<\psi_z(\omega)|\psi_z(\omega)\>}
\eeq
where $\vec\sigma$ are the Pauli
matrices. Note that these vectors are normalized, $|\vec m_z(\omega)|=1$, indicating that
individual frequency components are always fully polarized. However, the polarization
of the photon, given by
\beq
\vec M_z=\int_0^\infty |\psi_z(\omega)|^2 \vec m_z(\omega) d\omega,
\label{Meq}
\eeq
can be partially ($|\vec M_z|<1$) or even totally ($|\vec M_z|=0$) depolarized. 

If fully polarized light, e.g. from a laser diode, is launched into a fiber (position
$z=0$) one has: $\vec m_0(\omega)=\vec m_{laser}$ for all $\omega$, hence 
the photons are totally polarized: $|\vec M_0|=|\vec m_{laser}|=1$. Let us model the
optical fiber as a concatenation of trunks of length $\ell_j$ and
birefringence $\vec\beta_j$ \cite{PMDGisinPellaux}. 
Accordingly (neglecting losses) the photon states evolves to:
\beq
\psi_\ell(\omega)=e^{i\omega\ell_n\vec\beta_n\vec\sigma/2}...e^{i\omega\ell_2\vec\beta_2\vec\sigma/2}
e^{i\omega\ell_1\vec\beta_1\vec\sigma/2}
\psi_0(\omega)
\eeq
where $\ell=\sum_j\ell_j$ is the total length of the fiber.
In long fibers the $\vec\beta_j$, in particular their orientations, are random
\cite{PMDGisinPellaux}, hence the output light is depolarized: $\vec M_\ell\approx 0$.

Depolarization, i.e. decoherence in the polarization domain, severely limits potential
applications of quantum cryptography \cite{PhysWorld98}. Indeed, coding the qubit in polarization becomes
clearly unpractical, while coding the qubit in the phase \cite{QCphase} 
is no better because phase
decoding requires interferometers and interferences are sensitive to polarization. One
possible way out is to limit the width of the optical spectrum, so that the integral
in (\ref{Meq}) is dominated by the central frequency and the output light remains polarized. But even
so, polarization fluctuations would impose active feedbacks. A more elegant and practical
solution exploit the feature of Faraday Mirrors (FM) \cite{FM}. A FM consists of a
$\lambda/4$ Faraday rotator followed by an ordinary mirror (with normal incidence). The
effect of such a FM is to turn any incoming polarization state to its orthogonal
state, as illustrated on figure 3. This non-unitary transformation is possible because
one uses a description in which one switches from a right handed reference frame before
the reflection to a left handed one after the reflection. This is quite convenient
(though not necessary), as doing so the polarization transformations during
propagation back up the fiber are precisely the inverse of those the photon underwent on the way
to the FM:
\beq
\psi_{2\ell}(\omega)=e^{-i\omega\ell_1\vec\beta_1\vec\sigma/2}e^{-i\omega\ell_2\vec\beta_2\vec\sigma/2}...
e^{-i\omega\ell_n\vec\beta_n\vec\sigma/2} T_{FM} \psi_\ell(\omega)
\eeq
where $T_{FM}$ denote the transformation due to the FM and $\psi_{2\ell}(\omega)$ is
the polarization state after a go-and-return through the fiber. The effect of this is
easier analyzed using the Poincar\'e vectors $\vec m_z(\omega)$ for which the FM transformation
simply reverses the orientation: $\tilde T_{FM}~\vec m = -\vec m$ (where $\tilde T_{FM}$ is the
corresponding $T_{FM}$ operator but acting on the Poincar\'e vectors). Accordingly, 
$\vec m_{2\ell}(\omega)=-\vec m_{laser}$ for all $\omega$, hence
$\vec M_{2\ell}=-\vec m_{laser}$ and the return light is again fully polarized. Moreover,
the state of polarization is fixed (relative to the source). This result holds as long as
the fiber can be considered as fixed during the time of a go-and-return, typically some
micro-seconds. That this is indeed the case for km long installed telecom fibers was 
first demonstrated in \cite{QCApplPhysLett}: more than 99.8\% of repolarization was achieved
on a 23 km long fiber below lake Geneva.

This way of "polarization recoherence" is exploited in our "Plug \& Play" implementation
of quantum cryptography \cite{QCApplPhysLett,QCElett97,QCElett98}, see figure 4.

\section{Chromatic Dispersion: Decoherence in the frequency domain}\label{CD}
For long distance Bell experiments, the use of polarization correlation is unpractical
because of the depolarization mechanism described in the previous section
(see however \cite{BellInnsbruck98} where the distance and the photon spectrum were reduced 
to limit depolarization). Moreover
the use of Faraday Mirrors is incompatible with the requirement that the two
detectors and the source should be at three widely separated locations. In 1989 Jim Franson
\cite{Franson89} proposed an elegant two-photon interferometer free of the 
depolarization problem and suitable for tests of the Bell inequality over long
distance, see figure 5 (actually, in this scheme polarization has to be
controlled inside the two distant interferometers, but depolarization in the long fibers
connecting the source and interferometers is irrelevant \cite{TittelEuroLett,TittelPRA99}).
However, chromatic dispersion, the fact that different optical frequencies (wavelengths) 
propagate at different speeds, imposes severe limitations to the fringe visibility in Franson
interferometers. Indeed, the two photons, emitted precisely at the same time by
spontaneous parametric downconversion in a nonlinear crystal, 
must be detected in coincidence, within a time
window of typically 300 ps. This time window must be short enough so that one can distinguish
the cases when the two photons took both the short or both the long arm of their
interferometer, from the cases when they took different arms. For stability reasons it is
reasonable to have arm length differences of some tens of cm, corresponding to a few ns.
But if chromatic dispersion (or any other cause of dispersion) reduces the time correlation
between the photons, then a coincidence detection no longer guaranties that both
photons took the same path. Hence chromatic dispersion severely reduces the 2-photon
interference visibility.

In a dispersive media, like the silica of optical fibers, the chromatic dispersion vanishes 
for a wavelength close to 1310 nm (the exact value depends on details of the manufacture).
In our long distance Bell experiment, the single photons had a spectral width of about
$\pm 35$ nm. Hence, for a fiber length of 17 km \footnote{In our experiment the analyzers
were separated by slightly more then 10 km, but the connecting fibers were quite longer.}, 
the chromatic dispersion is of the order
of 500 ??? ps, large enough to reduce the fringe visibility down below the threshold set
by Bell inequality (Bell inequality is violated for visibilities larger than $1/\sqrt{2}\approx
71\%$). One way around this decoherence mechanism is the following. In good approximation
chromatic dispersion is a linear function of the wavelength
$\lambda$ (this approximation is valid over
several tens of nm). Hence the differential group delay is a quadratic function of
$\lambda$, with its minimum at $\lambda_0$, the wavelength of zero chromatic dispersion. 
Accordingly, if the central wavelength of the photon pair is precisely at 
$\lambda_0$, when, thanks to the frequency correlation of the two photons, both
photons are at wavelengths symmetrically above and below $\lambda_0$. Both photons
undergo thus the same differential group delay, hence arrive at the analyzer in perfect
coincidence. This phenomenon, called 2-photon chromatic dispersion cancellation
\cite{CD2photon}, is essential for long distance Bell experiments using optical fibers.
Note that we made two approximations: first that chromatic dispersion is approximately
a linear function of wavelength, next that the the frequency correlation $\nu_1+\nu_2
=\nu_{pump}$, due to energy conservation, implies the approximate wavelength
correlation $\lambda_1+\lambda_2\approx\lambda_{pump}$ (this second approximation is
not necessary, as all the discourse could be phrased in terms of frequency, but traditionally
chromatic dispersion is expressed in wavelengths). These approximations are in excellent
agreement with our experimental results \cite{TittelPRA97,TittelPRL98,TittelPRA99}.

\section{Conclusion}\label{conc}
Decoherence affects already systems of one and two-photon, setting limits to the
viability of the corresponding communication protocols. However, we have
illustrated how one can deal with decoherence for these relatively simple systems.
The generalization to larger, more complex, systems is not straightforward. Nevertheless,
there is hope that some of the ideas presented in this contribution may guide the 
research for improved "decoherence management".

\small
\section{Acknowledgments}
This work was partially supported by the Swiss National Science Foundation and by
the European TMR Network "The Physics of Quantum Information" through the Swiss
OFES.

\newpage
\normalsize
\section{Figure Captions}
\begin{enumerate}

\item Schematic of an instrument for Polarization Mode Dispersion measurements
using the so-called "interferometric method": the interferometer re-coheres photons
that went out of decoherence due to the polarization dispersion in the fiber under test.

\item Typical result obtained with the interferometric measurement method (see figure 1)
      for a standard telecom fiber.

\item Poincar\'e sphere picture of the polarization state transformation of a photon 
       when reflected by a Faraday mirror. Each state P undergoes first a 1/4 tour around
       the vertical axis, next a reflection due to the standard mirror, finally a second
       1/4 tour. As both rotations are due to the Faraday effect, they rotate in the same
       direction, despite that the photons are travelling in opposite directions. The final
       state P' is always opposite to the initial state P, i.e. P and P' represent orthogonal
       states.

\item Schematic of our "Plug \& Play" quantum cryptography system. For details see
       \cite{QCApplPhysLett,QCElett97,QCElett98}.

\item Principle of a Franson test of Bell inequality.

\end{enumerate}

\end{document}